\documentclass[prb,twocolumn,aps,showpacs,fixfloats]{revtex4}
\usepackage{graphicx}
\usepackage{bm}
\usepackage{amsmath,amssymb}
\usepackage{subfigure}
\usepackage{float}
\usepackage{latexsym}
\usepackage{epstopdf}
\usepackage{color}
\usepackage{enumerate}
\usepackage{pdfpages}

\begin{document}
\newcommand{\s}{\scriptscriptstyle}
\newcommand{\uu}{\uparrow \uparrow}
\newcommand{\ud}{\uparrow \downarrow}
\newcommand{\du}{\downarrow \uparrow}
\newcommand{\dd}{\downarrow \downarrow}
\newcommand{\ket}[1] { \left|{#1}\right> }
\newcommand{\bra}[1] { \left<{#1}\right| }
\newcommand{\bracket}[2] {\left< \left. {#1} \right| {#2} \right>}
\newcommand{\vc}[1] {\ensuremath {\bm {#1}}}
\newcommand{\tr}{\text{Tr}}
\newcommand{\Trans}{\ensuremath \Upsilon}
\newcommand{\Refl}{\ensuremath \mathcal{R}}


\title{Giant fluctuations of local magnetoresistance of organic spin valves and  non-hermitian 1D Anderson model}

\author{R. C. Roundy$^1$, D. Nemirovsky$^2$, V. Kagalovsky$^2$,  and M. E. Raikh$^1$} \affiliation{$^1$ Department of Physics and
Astronomy, University of Utah, Salt Lake City, UT 84112, USA\\
$^2$ Sami Shamoon College of Engineering, Beer-Sheva, 84100 Israel}

\begin{abstract}
Motivated by recent experiments, where the tunnel magnetoresitance (TMR) of a spin valve was measured {\em locally}, we theoretically study the distribution of TMR along the surface of magnetized electrodes. We show that, even in the absence of interfacial effects (like hybridization due to donor and
acceptor molecules), this distribution is very broad, and the portion of area with
negative TMR is appreciable even if {\em on average} the TMR is positive. The
origin of the local sign reversal is quantum interference of subsequent spin-rotation {\em amplitudes} in course of {\em incoherent} transport of carriers between the source and the drain.   We find the distribution of local TMR exactly by drawing upon
formal similarity between evolution of spinors in time and of reflection coefficient along a 1D chain in the Anderson model.
The results obtained are confirmed by the numerical simulations.
\end{abstract}

\maketitle

\begin{figure}[h!]
\includegraphics[width=77mm,clip]{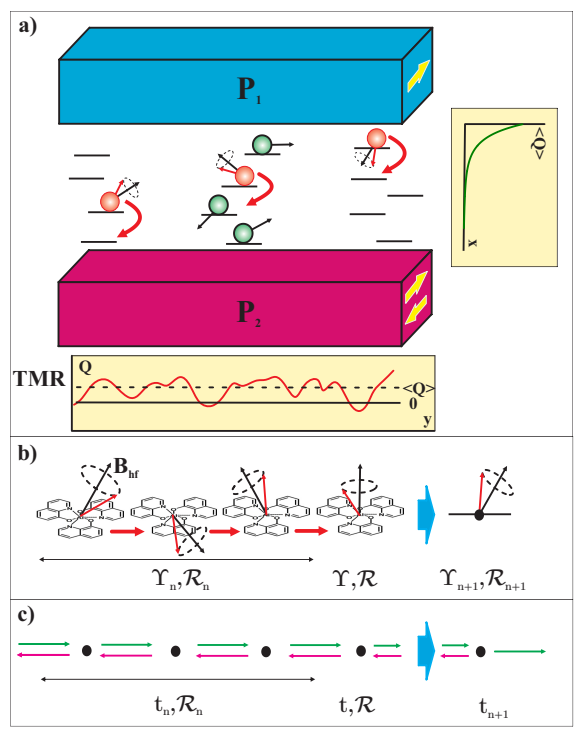}
\caption{(Color online) (a).  Schematic illustration of an OSV with a thin
active layer, so that the transport is along independent chains.
Electrode polarizations, $P_{1,2}$, are indicated in yellow.
The in-plane components of hyperfine fields are
depicted with black arrows.
Below: A cartoon of {\em local} TMR along
the $y$-direction; the classical value is indicated with a dashed line.
Right: Decay of the {\em average}
polarization across the active layer is shown.
(b) and (c).  Illustration of the mapping of {\em temporal} spin evolution
in course of hopping onto the {\em spatial} propagation of an
electron through a chain of random scatterers.
}
\label{analogy}
\end{figure}

\noindent {\em Introduction.} Organic spin valves (OSVs), being one of the most promising applications of organic spintronics,  are actively studied experimentally\cite{valve1,valve2,valve3,valve4,valve5,valve6,valve7,noHanle1,noHanle2}.
The organic active layer of an OSV is sandwiched between two magnetized electrodes.
Due to long spin-relaxation times of carriers in organic materials, the net resistance of OSV is sensitive to the relative magnetizations of the electrodes.
Among many advantages that OSVs offer, is wide
tunability  due to e.g. chemical doping, and enormous flexibility.
The processes that limit the performance of OSVs can be
conventionally divided into two
groups: (i) interfacial, which take place
at the interfaces between the electrodes and active layer \cite{Fertlocal,Nanolettlocal,chinese, negativeTMR1, negativeTMR2, negativeTMRinterface1, negativeTMRinterface2, negativeTMRimpurity},
and
(ii) intralayer,  which exist even if the interfaces are ideal.\cite{Bobbert, Flatte}
Due to the latter processes
the injected polarized  electrons, Fig. \ref{analogy}, lose memory about their initial spin orientation while traveling between the electrodes. One of the most prominent
mechanisms of this spin-memory loss is the precession of a carrier spin in random hyperfine fields of hydrogen nuclei\cite{valve5,Bobbert, Flatte}.
The effectiveness of the OSV
performance is
quantified by tunnel magnetoresistance (TMR)
given by a so-called modified Julliere's formula\cite{julliere}, see e.g. the review Ref. \onlinecite{PramanikReview},
\begin{equation}
\label{modified}
\text{TMR}=  \frac{2 P_1 P_2 \exp(-d/\lambda_s)}{1 - P_1 P_2 \exp(-d/\lambda_s)} ,
\end{equation}
\begin{figure}[t]
\includegraphics[width=97mm,clip]{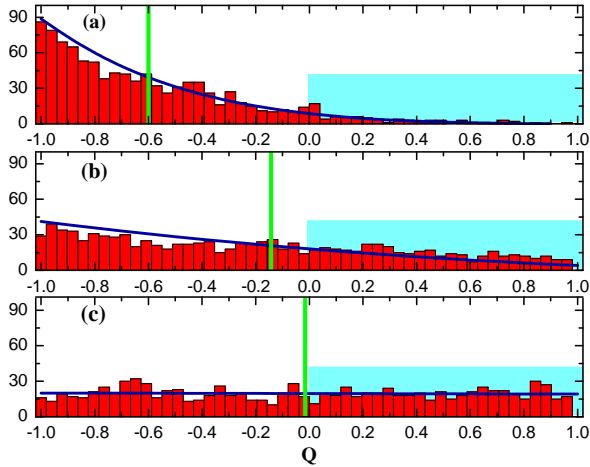}
\caption{(Color online).  Solid lines denote the distribution function
of the local degree of polarization, $Q$, for different rotation strengths on
the sites plotted from Eq. (\ref{solution}) for values $x = n\Refl^2 = 0.20 (a), 0.58 (b),$ and
$2.45 (c)$.  Red bins are the result of numerical
simulation of the system for $10^4$ random realizations of hyperfine fields,
with $n=20$ and $n\Refl^2$ as above. Classical values of polarization are shown
with green bars.  Blue rectangles highlight the domains of negative TMR.
 }
\label{fits}
\end{figure}
\begin{figure}[!ht]
\includegraphics[width=97mm]{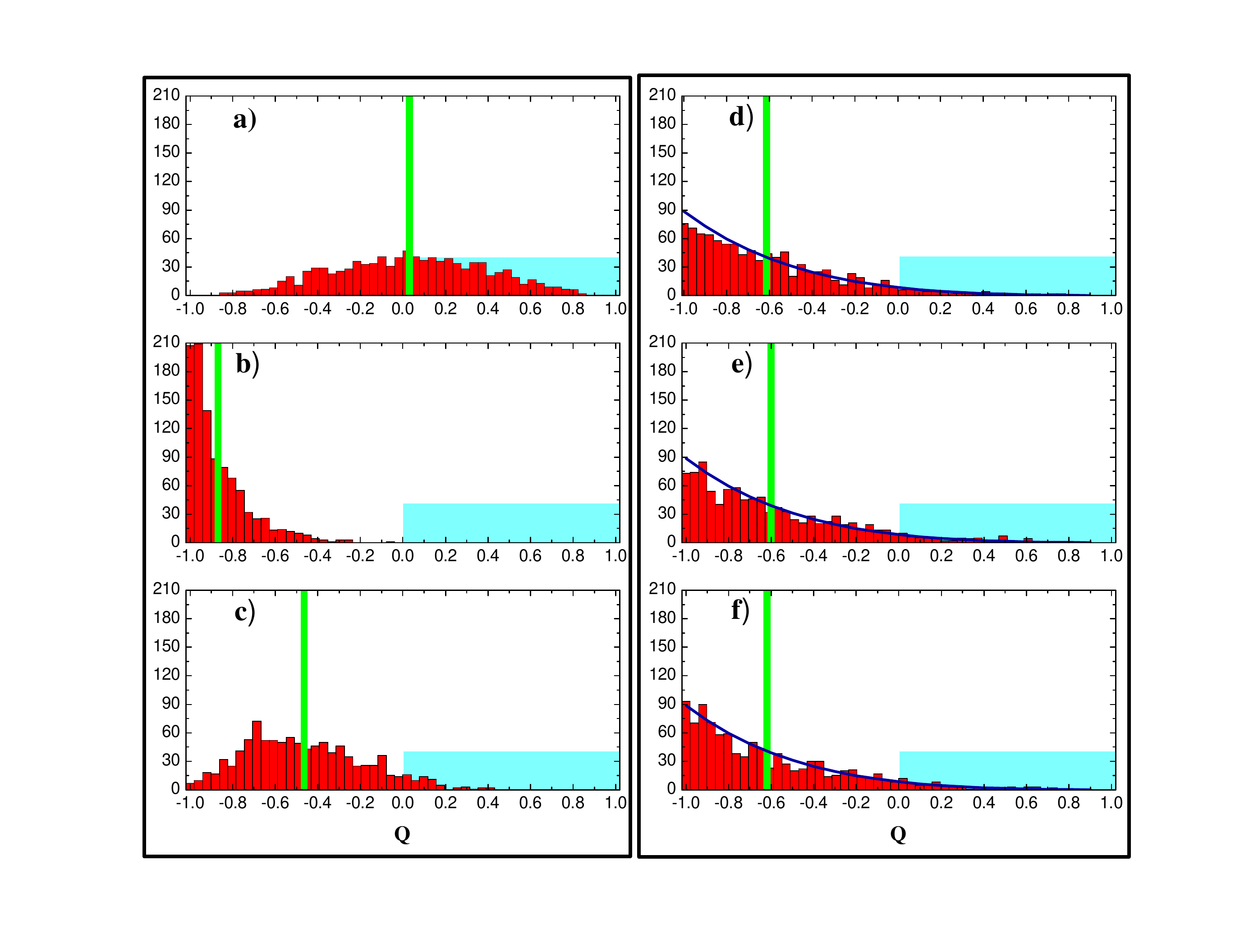}
\caption{(Color online).
Histograms of the spin polarization, $Q$, from simulation of the
system with  $n=20$, $\langle \Refl^2 \rangle = 0.01$.
 Left panel: In each of three histograms the orientations of hyperfine fields are {\em fixed}; the spread of $Q$-values is due to the randomness in the waiting times.
Right panel: Three histograms generated for the same number ($10^3$)
of realizations as in the left, but with allowance for randomness in the hyperfine-field orientations.
Values of TMR in the left panel, which are the averages of the histograms (green), are
specific for the configuration of the hyperfine field.  On the
contrary, the histograms in the right panel approach the theoretical
result, Eq. (\ref{solution}), shown with solid line.  These histograms
would represent the evolution of local TMR when field configuration
slowly rotates due to, e.g., spin-spin interaction.
%
%
%
%
%
}
\label{r-rand}
\end{figure}
where $P_1$, $P_2$ stand for polarizations of the electrodes. The difference from the original
Julliere's formula\cite{julliere} is the exponential factor $Q=\exp(-d/\lambda_s)$ describing the spin-memory loss over the active layer of thickness, $d$.
Processes (i) can be incorporated into Eq. (\ref{modified})
by appropriately modifying $P_1$, $P_2$. For example, in Ref. \onlinecite{Fertlocal} replacement of $P_1$, $P_2$ by ``effective'' spin polarizations reflects the relative position of the Fermi level with
respect to interfacial donor (acceptor) level. In this way, the ``effective'' polarization depends on bias, which might explain the sign reversal of TMR \cite{Fertlocal,Nanolettlocal,chinese, negativeTMR1, negativeTMR2, negativeTMRinterface1, negativeTMRinterface2, negativeTMRimpurity}.
Processes (ii), on the other hand, are
reflected in Eq. (\ref{modified}) via the factor $Q=\exp(-d/\lambda_s)$, where $\lambda_s$ is the spin diffusion length. The meaning of $Q$ is the polarization of electrons at $x=d$, provided that at $x=0$ they are fully polarized.
Encoding the processes (ii) into $Q=\exp(-d/\lambda_s)$ implies that the spin polarization of electrons falls off {\em homogeneously and monotonically} with coordinate $x$, see Fig. \ref{analogy}.

The prime message of the present paper is that strong  local fluctuations of TMR, including the local sign reversal, is a generic property of the  OSV even with ideal interfaces. In other words, the factor $Q$,
captures the spin memory loss  only {\em on average}. The {\em local} value of $Q$ fluctuates strongly from point to point and takes values in the domain $-1<Q<1$ .
On the physical level, the local value of TMR in the absence of interfacial effects, is the fingerprint of hyperfine-field configuration along a given current path.

The origin of strong local fluctuations of TMR is quantum-mechanical interference
of {\em amplitudes}\cite{our} of subsequent spin rotations accompanying the inelastic hops of
the electron
which has been routinely neglected in earlier studies. Formally, this interference,
in course of a {\em time} evolution of spin in random hyperfine field can be mapped on
{\em spatial} propagation of electron along a 1D disordered chain.\cite{berezinskii,pustilnik,withThouless,Izrailev}
In this regard, it is important to realize that, as electron enters the OSV, its spatial
coherence is lost after a single inelastic hop. At the same time, the spin evolution of a given electron remains absolutely coherent
all the way between the electrodes.

Experimental relevance of local TMR, which motivated our study, was demonstrated in a recent paper
Ref. \onlinecite{Nanolettlocal}, where an STM tip played  a
role of one of magnetized electrodes, while the other electrode
was a Cr(001) substrate with alternating magnetization directions.
The role of active layer was played by isolated $C_{60}$ molecules
attached to the substrate.  By scanning the tip, the authors were able to recover the surface map of the conductance through a single molecule, and its evolution with bias. In this way, the sign reversal of TMR was demonstrated on the {\em local} level.

\noindent {\em Recurrence relation for the spin transport.}
We will illustrate our message using the simplest model\cite{Bobbert,Flatte,multisteps,our} depicted in Fig. \ref{analogy}. As shown in this figure, electron hops along the parallel chains. The waiting times, $\tau_n$, for each subsequent hop are Poisson-distributed as
$\frac{1}{\tau^{\ast}}\exp[-\tau/\tau^{\ast}]$. While residing on a site, electron spin precesses around a
local hyperfine field.
Hyperfine fields are random,
their gaussian distribution
is characterized by rms value, $b_0$.

In course of hopping, the values $b_{\perp}$ change abruptly after each time interval,
$\tau_n$. The evolution of the amplitudes $a_1,a_2$  of $\uparrow$ and $\downarrow$ spin projections is described by the unitary evolution matrix defined as
\begin{equation}
\label{transfer}
\!\left\{\begin{matrix}a_1(\tau_{n+1}\!) \\ a_2(\tau_{n+1}\!) \end{matrix} \!\right\}\!=\!\widehat{U}_n \!\left\{\begin{matrix} a_1(\tau_n)\\ a_2(\tau_n)\end{matrix}\right\}\!,
~\!\widehat{U}
\!=\! \begin{bmatrix}
\Trans e^{-i \chi} \!&\! -i \Refl e^{-i \phi} \\
-i\Refl e^{i \phi} \!&\! \Trans e^{i \chi}
\!\end{bmatrix}
\!.
\end{equation}
Microscopic expressions for $\Refl$, $\Trans$, and the phases $\chi$ and $\phi$ are elementary:
$\Refl = \frac{|b_{n, \perp}|\tau_n}{2}$,
$\Trans = \sqrt{1-\Refl^2}$,
$\chi = \frac{b_z\tau_n}{2}$, and $\phi = \tan^{-1} \frac{b_{n,y}}{b_{n,x}}$. Here $b_z$ and $b_{\perp}=\left(b_x,b_y\right)$ are the tangetial and normal (with respect to the initial spin orientation) components of the hyperfine field.

Coherent evolution of the electron spin over $n$ steps
is described
by the
product, $\prod\limits_{i=0}^{n}\widehat{U}_i$,  of matrices Eq.~(\ref{transfer}). Naturally, after $n$ steps, this product can also be reduced to the form Eq. (\ref{transfer}) with $\Trans$ replaced by some effective $\Trans_n$. This observation suggests that  $\Trans_n$ and $\Trans_{n+1}$ are related via a recurrence relation which we choose to cast into the form
%
%
\begin{widetext}
\begin{equation}
\label{recurrence}
\frac{1}{\Trans_{n+1}^2}
= \dfrac{\dfrac{1}{\Trans_n^2}\dfrac{1}{\Trans^2}}{
1 + \left( \dfrac{1}{\Trans^2} - 1\right)
    \left( \dfrac{1}{\Trans_{n}^2} - 1 \right)
-  2\sqrt{
 \left( \dfrac{1}{\Trans^2}-1 \right)
 \left( \dfrac{1}{\Trans_n^2}-1 \right) }
 \cos\left( (-\phi_n\!+\!\phi)\!-\!(\chi_n\!+\!\chi) \right)
}.
\end{equation}
\end{widetext}

\noindent {\em Mapping on a 1D Anderson model.}
Consider now a different physical situation: {\em spinless} electron propagates
{\em coherently} along a line of impurities randomly positioned at points, $x_n$, see Fig. \ref{analogy}c. As shown in the figure, the energy-conserving wave function
on the interval $(x_n,x_{n+1})$ is a combination of two counterpropagating waves.
Denote with $t$ the amplitude transmission coefficient of the impurity. Then the relation between the net transmission coefficient, $t_n$, of the system with $n$ impurities satisfies the famous Fabry-Perrot-like recurrence relation
\begin{equation}
\label{localized}
t_{n+1}^2 = \frac{t_n^2 t^2}{1 + (1-t^2)(1-t_n^2) + 2 \sqrt{(1-t^2)(1-t_n^2)}\cos 2 \eta},
\end{equation}
where $\eta$ is the phase accumulation upon passage through the interval $(x_n, x_{n+1})$.

At this point we make our main observation that the recurrence relations
Eq. (\ref{localized}) and Eq. (\ref{recurrence}) map onto each other upon
replacement $\Trans^{-1} \leftrightarrow t$. On the other hand, it
is known that the distribution function of $t_n$ can be found exactly. In particular,
the average $\ln(t_n^2)$  increases linearly\cite{withThouless} with $n$, which is
the manifestation of the Anderson localization in 1D. Anderson localization is the result of quantum interference of multiply-scattered waves\cite{berezinskii}.
The very existence of the mapping of Eq. (\ref{recurrence}) onto
Eq. (\ref{localized}) suggests that the interference effects are equally important for the {\em temporal} evolution of spin. We will see, however, that replacement $t_n$ by $1/\Trans_n$ rules out the Anderson localization but causes giant fluctuations of $\Trans_n$ with $n$.  In addition, the mapping allows one to employ
well-developed techniques, see e.g. the review Ref. \onlinecite{Izrailev}, to describe these fluctuations analytically.

\noindent {\em Distribution of local spin polarization after $n$ steps.}
In the mapping of Eq. (\ref{recurrence}) onto
Eq. (\ref{localized}) the randomness of the impurity positions, $x_n$, is taken over by
the random azimuthal orientations of the hyperfine fields. Following Ref. \onlinecite{withThouless}, this randomness allows one to write down the  recurrence relation for the distribution function of the effective transmission coefficient. In our case it is more convenient to analyze the distribution of the related quantity $Q=2\Trans^2-1 = 1 - 2\Refl^2$,
which is the local spin polarization, as mentioned in the Introduction. Then the functional recurrence relation reads
%
%

\begin{widetext}
\begin{align}
\label{F-recursion}
{\cal F}_{n+1}(Q_{n+1}) &=
\int\limits_{-1}^1 dQ_n {\cal F}_n(Q_n)
\int\limits_0^{2\pi} \frac{d\phi}{2\pi}
 \delta \left[Q_{n+1} -  Q_n ( 1 - 2 \Refl^2) + 2\sqrt{1-Q_n^2} \Refl\Trans \cos\phi
 \right] \\
\label{F-recursion1}
 &= \frac{1}{\pi} \int\limits_{-1}^1 dQ_n
 \frac{{\cal F}_n(Q_n)}{
\sqrt{ \left[ 2\sqrt{1-Q_n^2} \Refl \Trans\right]^2 - \left[ Q_{n + 1} - Q_n (1-2 \Refl^2) \right]^2 }}. \nonumber
\end{align}
\end{widetext}
An immediate consequence of Eq. (\ref{F-recursion}) is the relation $\langle Q_{n+1}\rangle=(1-2\Refl^2)\langle Q_n \rangle$ between the averages.
This, in turn, implies that,
on average,
spin-memory loss follows the
classical prediction $\exp(-n/\lambda_s)$, where
$\lambda_s = 1/\ln(1-2\Refl^2)$.

From now on we consider the limit of large $n$ and small $\Refl$.  The latter allows
us to expand the denominator in Eq. (\ref{F-recursion}) to the first order
in $\Refl^2$, which, upon integration by parts, yields the following
Fokker-Planck equation
\begin{equation}
\label{F-P}
\frac{\partial{\cal F}}{\partial x} =
\frac{\partial}{\partial Q}\left[\left( 1-Q^2 \right) \frac{\partial {\cal F}}{\partial Q}\right],
\end{equation}
where $x=n\Refl^2$ is assumed to be a continuous variable.
It is not surprising that Eq. (\ref{F-P}) is {\em exactly} the Fokker-Planck equation
for 1D localization. The important difference, however, is that for spin evolution it
should be solved in the domain $|Q|<1$ rather than\cite{Izrailev} $Q>1$.
The latter is a direct consequence of the mapping $t\leftrightarrow \Trans^{-1}$.


For a restricted domain $|Q|<1$ the separation of variables in Eq. (\ref{F-P}) reveals that
the eigenfunctions  with respect to $Q$ are the Legendre polynomials, $P_m(Q)$, the corresponding eigenvalues, $m(m+1)$, define the $x$-dependence, $\exp[-m(m+1)x]$, for a given $m$. The coefficients in the linear combination of the Legendre polynomials are fixed by
the ``initial" condition ${\cal F}(0,Q)=\delta (Q+1)$, which corresponds to a full polarization at $x=0$. This yields the following solution ${\cal F}(x,Q)$
\begin{equation}
\label{Legendre}
{\cal F}(x, Q) = \sum\limits_{m=0}^{\infty} \left( m + \frac{1}{2} \right)
(-1)^{m} P_{m}(Q) e^{ -m(m+1) x }.
\end{equation}
Summation over $m$ in Eq. (\ref{Legendre}) can be performed explicitly by using
the integral presentation
\begin{equation}
e^{-m^2 x} = \int \frac{d\kappa}{\sqrt{\pi x}} \exp\left(
-\frac{\kappa^2}{x} + 2 i m \kappa
\right),
\end{equation}
and the identity
\begin{equation}
\label{identity}
\sum\limits_{m=0}^{\infty} \left( 2m + 1 \right) P_{m}(Q) \zeta^m
 =  \frac{1 - \zeta^2}{(1-2Q\zeta + \zeta^2)^{3/2}},
\end{equation}
which can be easily derived from the generating function for the Legendre polynomials.
Substituting $\zeta = -\exp(x - 2 i\kappa)$ and integrating by parts  leads to the final result
\begin{widetext}
\begin{equation}
\label{pre-solution}
{\cal F}(x,Q) = \frac{e^{x/2}}{i \sqrt{\pi x^3}}
\int dk (k+ix)
\frac{\exp\left( -\frac{k^2}{4x}-\frac{ik}{2} \right)}{
\sqrt{2Q + e^{x-ik} + e^{-(x-ik)}}}.
\end{equation}
The imaginary part of the integrand is odd in $k$. Therefore,
we can ultimately present Eq. (\ref{pre-solution}) as a purely
real integral
\begin{multline}
\label{solution}
{\cal F}(x,Q) = \frac{e^{x/2}}{4 \sqrt{\pi x^3}}
\int\limits_0^\infty \frac{dk \, \exp\left(-k^2/4x\right)}{
\left\{
\left[ (\cos k + Q \cosh x)^2 + (1-Q^2) \sinh^2 x \right]^{1/2}
+\left[
 e^{-x}\cos^2 k+Q\cos k +\sinh x
 \right]
\right\}^{1/2}}\\
\times \left(
x - \frac{\left(e^{-x}\cos k+Q\right)\left(k\sin k -x \cos k\right)-x\sinh x
 }
 { \left[  (\cos k + Q \cosh x)^2 + (1-Q^2)\sinh^2 x \right]^{1/2} }
 \right).
\end{multline}
\end{widetext}

The difference between Eq. (\ref{solution}) and its counterpart\cite{Izrailev} in 1D Anderson model
stems from the fact that the denominator in the identity Eq. (\ref{identity}) in our case is complex.

%
%
%

\noindent {\em Numerical results and analysis.}
The parameter $x=n\Refl^2$ in the argument of the distribution
Eq. (\ref{solution}) is related to the sample thickness, $d$ and
{\em classical} spin-diffusion length as $x=d/2\lambda_s$.
It is seen from Fig. \ref{fits}
 that, as $x$ passes through $x\sim 1$, the distribution
evolves from $\delta$-function (at $x\ll 1$) to linear and, eventually, to flat.
Flat distribution manifests  complete spin-memory loss. But even
when this loss is small on average, a sizable part of the distribution lies in the domain  $Q>0$, which corresponds to negative TMR. Note that, upon neglecting interference in Eq. (\ref{F-recursion}), the
distribution becomes $\delta\left(Q+e^{-2x}\right)$, i.e. infinitely narrow.

Until now we neglected the
effects  caused by the randomness of the waiting times, $\tau_i$. With regard to the distribution ${\cal F}(x,Q)$, this randomness amounts to replacement of $\Refl^2$ by
$\langle \Refl^2 \rangle_{\tau_i}$ in the parameter $x$. A much more delicate issue
is whether or not the randomness in $\tau_i$ affects the local value of TMR. Naturally, the TMR, measured by a local probe, is the average over all
$\tau_i$.
Then the question arises whether  this averaging washes out the difference between
the points at which the TMR is measured, i.e. replaces the local $Q$ by $\exp(-d/\lambda_s)$ or, on the contrary,
the averaged TMR is a unique signature of the actual realization
of the hyperfine fields along a given current path. We argue that
the second scenario holds. Our argument is two-fold. Firstly, we performed direct numerical simulation of local spin polarization along a {\em given} path with randomness in $\tau_i$ incorporated\cite{Supplement}.
The results shown in Fig. \ref{r-rand} demonstrate that, while
this randomness broadens the histograms, their center,
which is the observable
quantity, depends dramatically on actual orientations of the hyperfine fields along the path.
Secondly, our analytical calculation\cite{Supplement} demonstrates that, while the disorder due to
random orientations is short-ranged, $\langle b_x(\tau) b_x(\tau') \rangle_{b_{i,x}} = \exp\left(-\frac{|\tau-\tau'|}{\tau^*}\right)$, the same correlator calculated with {\em given}
hyperfine-field realization but with random $\tau_i$ falls off very slowly, as a power law. On the basis of two preceding arguments we conclude that, at times scales where nuclear spin-spin interaction
does not rearrange the hyperfine-field configuration, the TMR remains specific for this configuration.

\noindent {\em Concluding remarks.}
Our theory applies for OSVs with thin inhomogeneous active layers, depicted in Fig. \ref{analogy}, in which the transport can be modeled with
directed non-crossing paths\cite{noHanle2}.

In this paper we treated the time evolution of the amplitudes $(a_1, a_2)$ in terms of a product of matrices.  An alternate approach would be to start from the
Schr{\"o}dinger equations, namely,
$i \dot{a}_1 = \frac{1}{2} b_\perp(\tau) a_2(\tau)$,
and
$i \dot{a}_2 = \frac{1}{2} b_\perp^*(\tau) a_1(\tau)$.
These two equations can be reduced to a single second-order equation for, say, $a_1$.
This equation can then be reduced to the Schr{\"o}dinger-like form. This procedure would
formally demonstrate why the spin evolution maps on {\em non-hermitian} 1D Anderson model: the effective
potential, $\frac{1}{2b_{\perp}^2}\left(\ddot{b}_\perp b_{\perp}
- \frac{3}{2}\dot{b}_{\perp}^2 \right)$,
 in the Schr{\"o}dinger equation appears to be {\em complex}\cite{pustilnik}.


\noindent {\em Acknowledgements.}
We are grateful to Z. V. Vardeny for motivating us.  Also we are
extremely grateful to V. V. Mkhitaryan for reading the manuscript and
very useful suggestions.
This work was supported by the NSF through MRSEC DMR-112125, and by the BSF Grant No. 2010030.

\begin{widetext}
\section{Supplemental material}

\subsection{Distribution of off-diagonal element of the evolution matrix}

The expression
$\Refl = \left|b_{n,\perp}\right| \tau_n/2$ for the off-diagonal element of the evolution matrix applies in the limit of weak rotation, $\Refl \ll 1$. The spread in the local values
of $\Refl$ originates from the randomness of $b_{n,\perp}=(b_x,b_y)$ as well as from the randomness of the waiting times, $\tau_n$.
Therefore the calculation of the distribution function of $\Refl$ involves  averaging over three random variables
\begin{equation}
\label{R-distribution}
H(\Refl) = \int \frac{d^2 b}{\pi b_0^2}\exp\left( -\frac{b_\perp^2}{b_0^2} \right)
\int\limits_0^\infty d\tau F(\tau)
\delta\left( \Refl - \frac{\left| b_\perp \right| \tau}{2} \right),
\end{equation}
where $F(\tau)=\frac{1}{\tau^*}\exp\left( -\frac{\tau}{\tau^*} \right)$ is the Poisson distribution.
Introducing dimensionless variables, $x=b/b_0$ and integrating
over $\tau$ with the help of the $\delta$-function yields
\begin{equation}
\label{above}
H(\Refl) = \frac{4}{b_0 \tau} \int\limits_0^\infty dx \; \exp\left(-x^2 -\frac{2 \Refl}{b_0 \tau^* x} \right).
\end{equation}
For $\Refl > b_0 \tau^*$ the above integral can be calculated using the steepest-descent
method
\begin{equation}
\label{steepest}
H(\Refl) = \frac{2}{\sqrt{3}}\left( \frac{2\sqrt{\pi}}{ b_0 \tau^*} \right) \exp\left[ -3\left(\frac{\Refl}{b_0 \tau^*}\right)^{2/3} \right].
\end{equation}
It turns out that Eq. (\ref{steepest}) provides an excellent approximation for
{\em all} values of $\Refl$.  For example,  for $\Refl = 0$ the difference between
the exact value and Eq. (\ref{steepest}) amounts to a factor $2/\sqrt{3}$. We checked numerically that, with the latter distribution,
the histograms of local polarization do not differ from box-like
distribution.

Another effect of randomness in the waiting times originates from
the phase $\chi=\frac{b_z\tau}{2}$ in the matrix Eq. (\ref{transfer}).
Thus a rigorous account of the spread in $\tau_i$ requires generating
random $\chi_i$ and $\Refl_i$ from the joint distribution
\begin{equation}
\label{joint}
\tilde{H}(\Refl, \chi) = \int \frac{d^3 b}{(\pi b_0^2)^{3/2}}\exp\left( -\frac{ b^2}{b_0^2} \right)
\int\limits_0^\infty d\tau F(\tau)
\delta\left( \Refl - \frac{\left| b_\perp \right| \tau}{2} \right)
\delta\left( \chi - \frac{\left| b_z \right| \tau}{2} \right).
\end{equation}
Since typical $\Refl$ and $\chi$ are of the same order, we again used in the simulations the $\Refl_i$-values uniformly distributed between $0$ and $\Refl$ and $\chi_i$ values uniformly distributed between $-\frac{\Refl}{2}$ and $\frac{\Refl}{2}$. The results are shown in Fig \ref{r-rand}.

%
%
%
%
%
%
%
%
%
%

\subsection{Temporal correlators of the random fields}
Consider a hopping chain  containing $N \gg 1$ sites.
For concreteness we will consider only the  correlation of the $x$-projections
of the hyperfine fields. In course of
transit between the electrodes, the carrier spin ``sees" this projection in the form of a telegraph signal
\begin{equation}
b_x(t)=\sum_{i=0}^N (b_{i+1}-b_{i})\theta(\tau-\sum_{j=0}^i \tau_j),
\end{equation}
where $\theta(\tau)$ is a step-function, $b_i$ is the $x$-projection on site $i$, and $\tau_i$ are the random waiting times for the hop $i \rightarrow (i+1)$.
As was mentioned in the main text, there are two correlators, $\langle b_x(\tau)b_x(\tau+T)\rangle$, relevant for TMR.
The first is
\begin{equation}
\label{K1}
K_1(T) = \Big< b_x(\tau)b_x(\tau+T)\Big>_{\left\{\tau_i\right\}},
\end{equation}
for a {\em fixed} realization, $\left\{b_i\right\}$, and randomness
 coming only from the Poisson distribution of $\tau_i$. The second correlator, $K_2$, is $K_1$ averaged
over all possible realizations of hyperfine fields
\begin{equation}
\label{K2}
K_2(T) = \Big< b_x(\tau)b_x(\tau+T)\Big>_{\left\{b_i\right\}, \left\{\tau_i\right\} }
 = \langle K_1(T) \rangle_{\left\{b_i\right\}}.
\end{equation}
It is easy to see that $K_2(T)$ has a simple form
\begin{equation}
\label{K2-result}
K_2(T) =  b_0^2\exp(-T/\tau^*),
\end{equation}
and decays on the time scale of a single hop $\sim \tau^*$.
On the other hand, as we will see below, $K_1(T)$ persists at much longer times.
The result, Eq. (\ref{K2-result}), can be established from the
simple reasoning: the product $b_x(\tau)b_x(\tau+T)$ contains the terms
of the type $b_i^2$ and the terms $b_ib_j$ with $j\neq i$. The latter terms
vanish upon configurational averaging. The terms $b_i^2$ are nonzero only if
$T$ is smaller than $\tau_i$. The corresponding probability can be expressed as $\theta(\tau_i-T)$.
Subsequent averaging over $\tau_i$ leads us to Eq. (\ref{K2-result}).

Turning to the correlator $K_1$,
in order to perform averaging over $\tau_i$ in Eq. (\ref{K1}) we use the integral representation of the $\theta$-function and cast $b_x(\tau)$ in the form
\begin{equation}
b_x(\tau) = \int \frac{d\omega \;e^{-i\omega \tau}}{-2 \pi i \omega} \left(
b_1 e^{i \omega \tau_1} + (b_2 - b_1) e^{i \omega (\tau_1 + \tau_2)} + \cdots
\right).
\end{equation}
In a similar way the product $b_x(\tau)b_x(\tau +T)$ can be presented as a double integral
\begin{multline}
\label{doublesum}
b_x(\tau)b_x(\tau+T) = \int \frac{d\omega \;e^{-i\omega \tau}}{-2 \pi i \omega}
\int \frac{d\omega' \;e^{-i\omega' (\tau+T)}}{-2 \pi i \omega'}
 \left(
b_1 e^{i \omega \tau_1} + (b_2 - b_1) e^{i \omega (\tau_1 + \tau_2)} + \cdots
\right) \\ \times
\left(
b_1 e^{i \omega' \tau_1} + (b_2 - b_1) e^{i \omega' (\tau_1 + \tau_2)} + \cdots
\right).
\end{multline}
The advantage of  the above representation is that
it allows averaging over $\tau_i$ in the integrand using the relation
%
\begin{equation}
\langle e^{i\omega \tau_i}\rangle =\int_0^{\infty}d\tau\,e^{i\omega \tau}F(\tau)= \frac{1}{1-i\omega\tau^*}.
\end{equation}
Obviously, the average $\langle b_x(\tau)b_x(\tau +T) \rangle$
does not depend on $\tau$, since it should be understood as
$\lim\limits_{\mathcal{T}\rightarrow \infty}\frac{1}{\mathcal{T}}\int_0^{\mathcal{T}}d\tau \langle b_x(\tau)b_x(\tau+T) \rangle$. Then the integration over $\tau$ sets $\omega=-\omega'$.

The coefficient in front of $b_i b_j$-term in the product  Eq. (\ref{doublesum}) is given by
\begin{equation}
\label{BigProduct}
\frac{\exp(-i \omega T)}{2 \pi \omega^2}
\left[
\exp\left(i \omega \sum\limits_{k=0}^{i} \tau_k\right)
- \exp\left(i \omega \sum\limits_{k=0}^{i-1} \tau_k\right)
\right]
\left[
\exp\left(-i \omega \sum\limits_{k=0}^{j} \tau_k\right)
- \exp\left(-i \omega \sum\limits_{k=0}^{j-1} \tau_k\right)
\right].
\end{equation}
Assume  that $j$ is smaller than $i$, then all terms with $k< (j-1)$
do not enter into Eq.  (\ref{BigProduct}).
As a result, the  averaging over remaining $i-j+1$ random times leads to the following result for the coefficient in front of $b_ib_j$

\begin{equation}
\dfrac{\exp(-i\omega T)}{2 \pi}
\left\{
\begin{matrix}
 \dfrac{1}{(1-i \omega \tau^*)^{i-j+1}},
& i \ge j \\
\dfrac{1}{(1+i \omega \tau^*)^{j-i+1}},
& i < j
\end{matrix}
\right. .
\end{equation}
The remaining step is the  integration over $\omega$ in Eq. (\ref{doublesum}). This integration
is carried out straightforwardly  by closing the contour in the bottom
half of the complex $\omega$-plane. The $\omega$-integral is nonzero only for $j \le i$.
The final expression for the coefficient in front of $b_i b_{j}$ reads
\begin{equation}
\int \frac{d\omega}{2\pi} \frac{e^{-i\omega T}}{(1-i\omega\tau^*)^{i-j+1}}
  = \frac{1}{(i-j)!}\left( \frac{T}{\tau^*} \right)^{(i-j)} \exp\left(-\frac{T}{\tau^*}\right).
\end{equation}
Thus the final result for the  correlator $K_1$ acquires the form
\begin{equation}
\label{K1-final}
K_1(T) = K_2(T)
+
\frac{1}{N} \sum\limits_{i=1}^{N} \sum\limits_{j<i} b_i b_{j} \left[ \frac{1}{(i-j)!}
\left( \frac{T}{\tau^*} \right)^{(i-j)} \exp\left( -\frac{T}{\tau^*} \right) \right].
\end{equation}
Note that if we perform averaging over $b_i$-s, the second term will vanish, and we will  recover the expected result
Eq. (\ref{K2-result}).
For non-averaged $K_1$ the term in the square brackets
restricts the domain of summation over $(i-j)$ to $\left| i-j-\frac{T}{\tau^*}\right|
\le \sqrt{\frac{T}{\tau^*}}$.
Therefore, if the length of the chain, $N$,
is smaller than $\frac{T}{\tau^*}$ the above condition will  {\em never} be
satisfied, and $K_1$ will fall off exponentially with $T$ with characteristic decay time
$\tau^*$.  In the opposite limit, the summation over $i,j$ within the
 allowed domain will eliminate $T$ dependence from $K_1(T)$.
 We can now restate
 the above observation as follows: $K_1(T)$ weakly depends on $T$ for $T<N \tau^*$, and
 decays exponentially with $T$ for $T>N \tau^*$.
 Since the transport of electron between the electrodes takes the time $N\tau^*$, we conclude that the realization of the hyperfine field does not change during this time interval.

Finally, to estimate the magnitude of $K_1(T)$ for $T<N\tau^*$, we calculate the quantity $K_1^2$ and average it over hyperfine fields.  This averaging can be performed analytically.  The result is conveniently expressed through the modified Bessel function, $I_0$, as
\begin{equation}
\label{rms}
\langle K_1(T)^2 \rangle_{\left\{ b_i \right\}} =  b_0^4 I_0\left( 2 \frac{T}{\tau^*} \right) \exp\left( -2\frac{T}{\tau^*}\right).
\end{equation}
For $T > \tau^*$ Eq. (\ref{rms}) simplifies to $K_1(T)^2 = b_0^4 / \sqrt{4 \pi T/\tau^*}$.

%
%
%
%
%
%
%

\subsection{Broadening of Classical Distribution}

As it was mentioned in the main text, neglecting the interference in Eq. (\ref{F-recursion}) leads to the infinitely sharp distribution,
${\tilde{\cal F}}(x, Q)=\delta (Q+e^{-2x})$. This conclusion however implies that
the magnitudes of  $\Refl$ are the same on each site. In reality the magnitudes of $\Refl$ are distributed according to Eq. (\ref{steepest}). This will
cause a broadening of the classical distribution function, ${\tilde{\cal F}}(x, Q)$,
which we estimate below. 

We begin with the recurrence relation for  ${\tilde{\cal F}_n}(Q)$
\begin{equation}
\label{recurrence-classical}
{\tilde{\cal F}}_{n+1}(Q_{n+1})
= \int\limits_{-1}^1 dQ_n \; {\tilde{\cal F}}_{n}(Q_{n})
  \int d\Refl \; H(\Refl) \; \delta\left( Q_{n+1} - Q_{n}(1-\Refl^2) \right).
\end{equation}
The explicit form of ${\tilde{\cal F}}_{n}(Q)$ can be found exactly 
for arbitrary distribution $H(\Refl)$. For this purpose we introduce a 
new variable $z=\ln Q$ and rewrite Eq. (\ref{recurrence-classical}) in 
terms of the function $G(z)=e^z{\tilde{\cal F}}(e^z)$,
\begin{equation}
\label{recurrence-classical-G} 
G_{n+1}(z) = \int d\Refl\; H(\Refl) G_{n}\left(z - \ln(1-2\Refl^2)\right).
\end{equation}
The right-hand-side of Eq. (\ref{recurrence-classical-G}) is a convolution and turns into a product upon the Fourier transform. This readily yields
\begin{equation}
G_{n}(k) = G_0(k) \left[ \int d\Refl \; H(\Refl) \exp\left( 2 i k \Refl^2 \right) \right]^n.
\end{equation} 
For large $n$ the form of $G_{n}(k)$ and, correspondingly, the form of the distribution $G_{n}(z)$ approaches to the Gaussian. Thus the distribution 
${\tilde {\cal F}}(Q)$ is essentially log-normal
\begin{equation}
\label{tilde}
{\tilde {\cal F}}_n(Q) = \frac{1}{|Q| \sqrt{\pi n \sigma_{\s \Refl^2}
}}
\exp \left[ \frac{\left(\ln |Q| + n \langle R^2 \rangle \right)^2}{
n \sigma_{\s \Refl^2}} \right],
\end{equation}
where $\sigma_{\s \Refl^2} = \left( \langle R^4 \rangle - \langle R^2 \rangle^2 \right)$. It follows from Eq. (\ref{tilde}) that the center of the distribution, ${\tilde {\cal F}}_n(Q)$, moves linearly with $n$, which is the same as the average of the quantum distribution,  while the width slowly grows with $n$ as $\delta Q=\sqrt{n\sigma_{\s \Refl^2}} \exp(n\langle\Refl^2\rangle)$.
Strong local {\em quantum} fluctuations of TMR persist up to $n\Refl^2 \lesssim 1$. For such $n$ the width of the classical distribution remains smaller than $\Refl$. Note also, that probabilistic treatment of the spin rotation encoded in Eq. (\ref{recurrence-classical}) forbids the negative TMR, i.e. restricts the domain of ${\tilde {\cal F}}_n(Q)$ to negative $Q$.

If we, however, proceed from the classical limit of Eq. (\ref{F-recursion}) to the Fokker-Planck equation, then the classical limit of the Fokker-Planck equation would correspond to neglecting $Q^2$ in the right-hand side of Eq. (\ref{F-P}).  Note that by doing so, we also remove the restriction that $Q$
is negative. The Fokker-Planck equation in this limit reduces to a heat equation, and, similarly to the quantum result, yields a flat distribution 
at large $x$.  This corresponds to ``temperature equilibration'' at long times.  Even though the classical and quantum Fokker-Plank results share 
limiting behavior and have the same average at all times, their shapes are
visibly distinct.

In this subsection we have demonstrated that there are two 
different classical limits of the quantum spin evolution. They predict
two dramatically different shapes for the distribution of spin polarization.

\end{widetext}
\end{document}